\documentstyle[11pt,newpasp,twoside,epsf]{article} 
\def\edcomment#1{\iffalse\marginpar{\raggedright\sl#1\/}\else\relax\fi} 
\reversemarginpar 

\begin{document} 
\title{A New Formation Mechanism for the Hottest Horizontal-Branch Stars}

\author{Allen V. Sweigart} 
\affil{NASA Goddard Space Flight Center, Code 681, Greenbelt, MD 20771}
\author{Thomas M. Brown} 
\affil{Space Telescope Science Institute, 3700 San Martin Drive,
Baltimore, MD 21218}
\author{Thierry Lanz, Wayne B. Landsman, Ivan Hubeny}
\affil{NASA Goddard Space Flight Center, Code 681, Greenbelt, MD 20771}

\begin{abstract} 
Stars with very large mass loss on the red-giant branch
can undergo the helium flash while descending the
white-dwarf cooling curve.  Under these conditions
the flash convection zone will mix the hydrogen
envelope with the hot helium-burning core.
Such ``flash-mixed'' stars will arrive on the
extreme horizontal branch (EHB) with helium- and carbon-rich
envelopes and will lie at higher temperatures than
the hottest canonical (i.e., unmixed) EHB stars.  Flash mixing
provides a new evolutionary channel for populating the
hot end of the EHB and may explain the origin
of the high gravity, helium-rich sdO and sdB stars.
\end{abstract}

\section{Evolution of Hot Flashers} 

Castellani \& Castellani (1993) first showed that stars
with large mass loss on the red-giant branch (RGB) can
evolve to high effective temperatures before igniting
helium in their cores.  We have studied the properties
of these ``hot flashers'' by following their evolution
through the helium flash to the zero-age horizontal branch (ZAHB)
for various amounts of mass loss (Brown et al. 2001;
Sweigart et al. 2002).  Two examples of our results are given
in Figure 1.  The ``early hot flasher'' in the left panel
ignites helium between the tip of the RGB
and the top of the white dwarf (WD) cooling curve.  Mixing
between the hydrogen envelope and helium core
does not occur in such stars due to the high
entropy of the hydrogen shell.  Early hot flashers will
thus have canonical, hydrogen-rich envelopes on the EHB.
    
The evolution is dramatically different for the ``late
hot flasher'' in the right panel of Figure 1.  Due to the weaker
hydrogen shell the flash convection zone will then penetrate
deeply into the stellar envelope.  Protons from the envelope
will be mixed into the helium core and rapidly burned, while
helium and carbon from the core will be mixed outward to the stellar
surface.  Such flash mixing is a consequence of the basic
properties of the stellar models and should therefore occur
whenever a star ignites helium on the WD cooling curve.
         
Flash mixing has many implications for the properties of the EHB stars:
     
\begin{itemize}
\item Flash mixing produces hot EHB stars with greatly enhanced
envelope helium and carbon abundances.
\item Flash-mixed EHB stars will appear subluminous in the UV
due to their redistributed far-UV flux and larger bolometric
corrections, in agreement with the subluminous EHB
stars observed in $\omega$~Cen and NGC~2808.
\item Flash mixing leads to a sharp dichotomy in the properties
of the EHB stars which may explain the gap within the EHB 
found in optical color-magnitude diagrams of NGC~2808.
\item Flash mixing on the WD cooling curve provides a new
evolutionary channel for producing hydrogen-deficient stars,
particularly the high gravity, helium-rich sdO stars and the
minority of sdB stars that are helium-rich.
\item The high carbon abundance in the envelopes of the
flash-mixed stars may be important for driving the pulsations
of the variable sdB stars.
\end{itemize}
             
\begin{figure*}[t]
\hspace{-0.30in}{\plotone{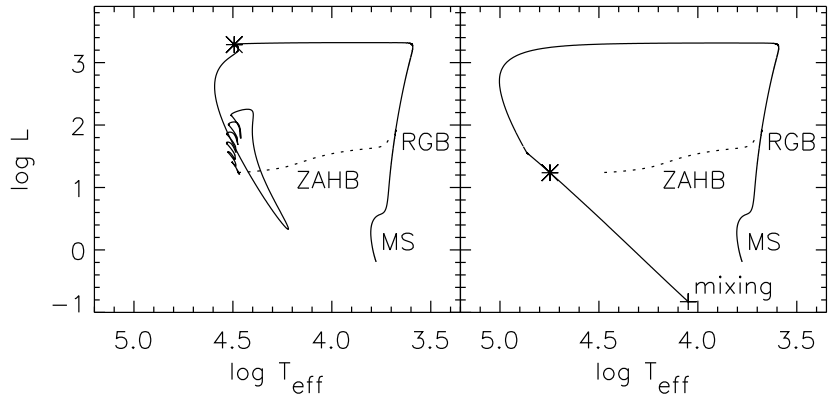}}
\caption{Evolution of an early hot flasher (left panel)
and a late hot flasher (right panel) from the main
sequence (MS) through the helium flash to the ZAHB.  The
flash peak is marked (*).  Flash mixing began
at the end of the late hot flasher track (+).}
\label{figure 1}
\end{figure*}


\begin{references}
\reference Brown, T. M., Sweigart, A. V., Lanz, T.,
Landsman, W. B., \& Hubeny, I. 2001, \apj, 562, 368
\reference Castellani, M., \& Castellani, V. 1993, \apj, 407, 649
\reference Sweigart, A. V., Brown, T. M., Lanz, T., Landsman,
W. B., \& Hubeny, I. 2002, in $\omega$~Centauri: A Unique Window into
Astrophysics, ed. F. van Leeuwen, J. D. Hughes, \& G. Piotto (San Francisco:
ASP), 261
\end{references}
\end{document}